\documentclass{article}
\usepackage{amssymb}
\usepackage{amsmath}
\usepackage{hyperref}
\usepackage{graphicx} % Required for inserting images
\usepackage{xr}
\usepackage{cleveref}
\usepackage{xcolor}
\usepackage{geometry}
\usepackage{subcaption}
\usepackage{xcolor}
\usepackage{authblk}
\usepackage{lineno}
\usepackage{xr}

\title{Energy pathway variety and the progress of the energy transition in European countries}

\author[1]{Noam Abadi\footnote{Corresponding author: n.abadi@rug.nl}}
\author[2,3]{Rossana Mastrandrea}
\author[1]{Franco Ruzzenenti}
\author[4]{Andrea Diaz}

\affil[1]{Energy and Environmental Research Institute of Groningen (ESRIG), Faculty of Science and Engineering, University of Groningen, Groningen, 9747 AG, the Netherlands}
\affil[2]{Department of Management, Faculty of Economics, University of Turin, Turin, Italy}
\affil[3]{IMT School for Advanced Studies, Piazza S. Francesco 19, Lucca, 55100, Italy}
\affil[4]{European Commission, Joint Research Centre (JRC), Seville, Spain}

\date{October 2025}

\begin{document}

%\linenumbers

\maketitle

\begin{abstract}
The integration of new energy forms into existing energy infrastructure has emerged as a critical challenge in the context of the pursuit of a sustainable energy transition. One of the main challenges is understanding how this integration takes place not only from the introduction, but also as energy follows existing paths or creates new ones through which it is transformed and used by different activities. Here we introduce techniques from network science to analyse this process for the case of 29 European countries between 1992 and 2021. We study how new energy forms increase or decrease the variety (heterogeneity) of paths through the system of each country by establishing new ones and replacing or phasing out existing ones. We find that the transition to systems based on renewable energy is characterised by an initial increase in the variety of paths while the heterogeneity of paths decreases at the end of the transition, when the proportion of non-renewables in the system tends to zero. We then demonstrate that greater heterogeneity (complexity) is associated with larger annual fluctuations in the proportion of non-renewable sources in the system, establishing a direct relationship between the progress of the transition and the complexity of the energy system in which it occurs.
This contributes to the understanding of general properties of the dynamics of the energy transition and effects that accelerate or deter it.
\end{abstract}

\section*{Integration of multi-energy systems, a review}

The supply, transformation and use of energy have an impact on the atmosphere \cite{liu2025increasing,paniw2022higher,emmerling2024multi} and the biosphere \cite{de2021adverse,liu2025drought,li2025extreme} that is difficult to understate. Emissions and pollutants originate from sectors such as transport, energy production, construction, manufacturing, and chemical processing, among others. Together, these activities play a central role in  shaping the structure and evolution of energy systems \cite{lamb2021review}. Energy systems involve the final use of energy in single industries, such as fuels for the transport or construction industries and electricity for the chemical industry or households, but also activities that extract and transform energy to supply it to final users in suitable forms; this is why they are sometimes referred to as \textit{multi-energy systems} \cite{mancarella2014mes}. 
The complexity and interdependence of activities and energy forms make the transition to a sustainable society particularly challenging \cite{bruckner2014energy,lopion2018review}. Moreover, the extent to which each energy form is embedded within specific infrastructures - connecting activities and transformations - largely determines the rate and feasibility of this transition \cite{smil2000perils}.

The study of energy system behaviour has garnered significant attention within energy systems modelling \cite{herbst2012introduction, pfenninger2014energy}, yielding important insights on how they can evolve to meet the dual objectives of sustainability and affordability \cite{bolwig2019review, bompard2022assessing, dangerman2013energy, beres2024impact}. Economics has played a key role in these studies, extending frameworks like input-output and supply-use analysis \cite{nakano2018development, pan2018dynamic, aramendia2022developing}, and integrating knowledge from environmental and social sciences (\cite{bazmi2011sustainable, fontes2018sustainable, pehl2017understanding}). Whereas these efforts have focused on optimizing costs, supply, and sustainability, questions concerning long-term dynamics \cite{fodstad2022next} - rather than real-time or inter-temporal actions - have been advanced through the application of System Dynamics and Agent-Based Modelling methodologies. These approaches enable the integration of both dynamic and stochastic features inherent to such systems.
Moreover, the growing recognition of systems as composed of interacting entities (such as individuals, artifacts, corporations, sectors) has attracted increasing attention from the field of complex network theory.  This discipline is grounded in the premise that the structure of connections between components of a system is fundamental to understanding its behaviour and emerging phenomena. While this is particularly intuitive in the context of energy systems, network theory offers formal frameworks and mathematical analytical tools to systematically capture such interdependencies  in different fields \cite{chu2017complex, sun2005complex, beyza2019applying, zhang2024identifying, zhang2024complex, chen2018energy}, moving  beyond conventional frameworks based on industry-to-industry or country/region/sector classifications \cite{hafner2020emergence, mir2020exploratory}.

From this perspective, several questions arise regarding the relationship between the energy transition and the intrinsic complexity of the energy systems. 
What is the nature of the relationship between an evolving energy system and its complexity? To what extent do the various components of the system interact with one another in shaping this transition? Is the current energy transition necessarily associated with increased complexity \cite{ruzzenenti2017complexification} with both intended and unintended consequences \cite{adekoya2023does,bakhsh2024strategy}, or do the energy imperatives driving this transition, as suggested by Valcav Smil \cite{smil2007energy}, serve as both a conduit and a condition for more complex systems? 
These questions also assume an answer to the most difficult and perhaps most enduring question of all, namely what is complexity and how should it be measured \cite{chaisson2002cosmic,morin1992concept}? 

In this study, we seek to address these questions by introducing a complex systems framework to analyse the energy systems of European countries over the period 1992–2021. We expand on the work of Diaz et al. (\cite{diaz2022understanding}) who, building on complex network theory, argued that the transition towards a cleaner use of energy adds complexity to the energy systems in terms of connectivity as their production and consumption patterns evolve. They explored the concept of sectoral integration within the energy systems of France and Luxembourg, demonstrating that, in the context of increasing electrification and an expanding share of renewable energy generation, these systems exhibited a discernible trend toward enhanced network connectivity.

Using their newly extended dataset  for the period 1992–2021 (JRC ENERNET dataset \cite{diaz2022energy}), we model the energy system of 29 European countries as a weighted directed network of activities and energy carriers (\textit{nodes}) connected through energy flows (\textit{links}).  Graph theoretical approaches based on the concept of \textit{random walks} allow to decompose energy flows into “effective transformation paths,” i.e., the fraction of energy that enters in one form and ultimately exits in another. Treating those path fractions as shares, we quantify system heterogeneity with the Herfindahl Index at country, sector group and single activity levels. 

The analysis reveals a strikingly uniform, parabolic trajectory: as the share of renewables increases, variety of paths, interpreted as structural complexity - initially rises, peaks at a common threshold of non-renewable share, and then declines. Greater variety also intensifies short-term fluctuations in the non-renewable share, suggesting that complex energy systems might become more vulnerable as they substitute non-renewables with renewables at a greater pace. These findings emphasize diversification, i.e. variety of paths, as a lever that could potentially accelerate and stabilize the European energy transition. We propose an economic mechanism to account for this behaviour and examine its implications for the broader energy transition.

\section*{European energy systems as networks}

Our analysis builds on the JRC-ENERNET database \cite{diaz2022energy}, which comprises 841 country-level networks, covering 29 European countries over a 29-year period, 1992–2021 (The Data subsection in Methods reports the list of countries and their acronyms used throughout the paper). The dataset translates the Eurostat’s energy balance statistics into networks of interconnected activities, with two categories of nodes: the first represents \textit{activities} related to the supply, transformation, and use of energy; the second represents \textit{energy carriers}, namely the physical forms in which energy is available (e.g., coal, petroleum, electricity). By construction, links can only occur from an activity node to a carrier node or from a carrier node to an activity node, but never between two nodes of the same type. These links are weighted according to the aggregated annual energy flow between the nodes, expressed in Tons of Oil Equivalent (TOE), and taking place in the physical form of the involved carrier node. For example, flows to or from the electricity carrier node will represent the yearly aggregated flows in the form of electricity, for example from production activities or to households, that takes place over the course of a year in a particular country.   
%The carrier nodes can be understood as the physical containers of each type of energy available in the system. For instance, the electricity node can be seen as a hub that gathers electricity from different production activities (electricity production nodes) and distributes it to other nodes where it is consumed, such as in demand sectors (e.g. buildings and transport). In this sense, the electricity node represents the transmission and distribution of electricity. This can be transposed to other carriers, such as liquid fuels (e.g. gasoline, diesel), where the node represents the distribution to storage facilities for final consumption. This has the advantage that it allows for capturing the flows of energy between activities in different forms without considering multiple-layered networks, each representing the flows of a particular energy form.

\begin{figure}[hbt!] %[ht!]
    \centering
    \includegraphics[width=0.6\linewidth]{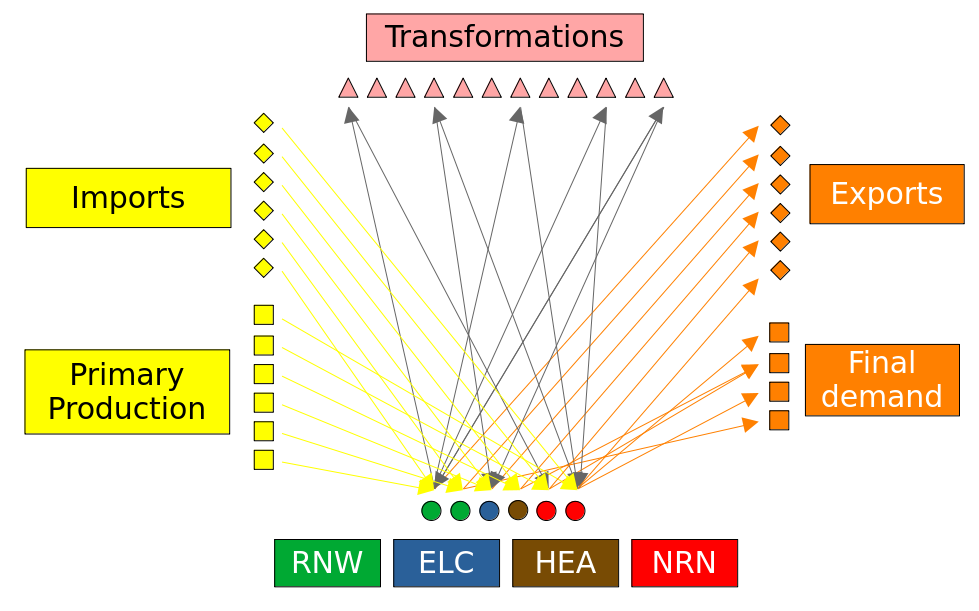}
    \caption{Network representation of a system with two types of nodes, carriers and activities. Flows only occur between carriers and activities, but not between carriers or between activities. Carriers can be non-renewable, renewable, electricity or heat. Activities that only have outgoing flows, shown on the left, are source activities (imports and primary production), while those that only have incoming flows, shown on the right, are sink activities (exports and final demand). In the middle, transformation activities and carriers have both incoming and outgoing flows.}
    \label{fig:network_representation}
\end{figure}

Furthermore, carrier nodes are classified into four groups: non-renewables (NRN), renewables (RNW), electricity (ELC) and heat (HEA). Likewise, activity nodes can be grouped into import (IMP) and primary production (PPRD) activities, which together we call \textit{source activities}, final demand (FD) and export (EXP) activities, referred to as \textit{sink activities}, and transformation activities (TA). These types are distinguished according to: (i)\textit{source a.}, with only out-going links; (ii) \textit{sink a.}, with only in-coming links; (iii) \textit{transformation a.}, with both types of connections. Additionally, import, primary production, and export activities are associated with a specific energy carrier, such that their flows originate from or terminate at a single carrier node. Figure \ref{fig:network_representation} shows the  energy system network with the described structure.

\section*{Connectivity and topological variation in European energy networks}

The 29 European countries analysed exhibit significant diversity in their energy system network structures, characterised by variations in network size (number of nodes) and connectivity (average degree). This diversity reflects differences in resource endowments, policy trajectories, and technological adoption. For instance, Germany, Austria, and France feature the most extensive networks (e.g., Germany had 147–164 nodes from 1992 to 2021), while Luxembourg, Iceland, and Cyprus maintain smaller, more compact systems (e.g., Cyprus had 46–80 nodes during the same period). Supplementary figures 1 and 2 (\cref{supp-fig:number_of_nodes,supp-fig:average_degree}) provide a detailed overview of how network size and average degree have evolved in all the countries studied.
\begin{figure}[ht!]
    \centering
    \includegraphics[width=\linewidth]{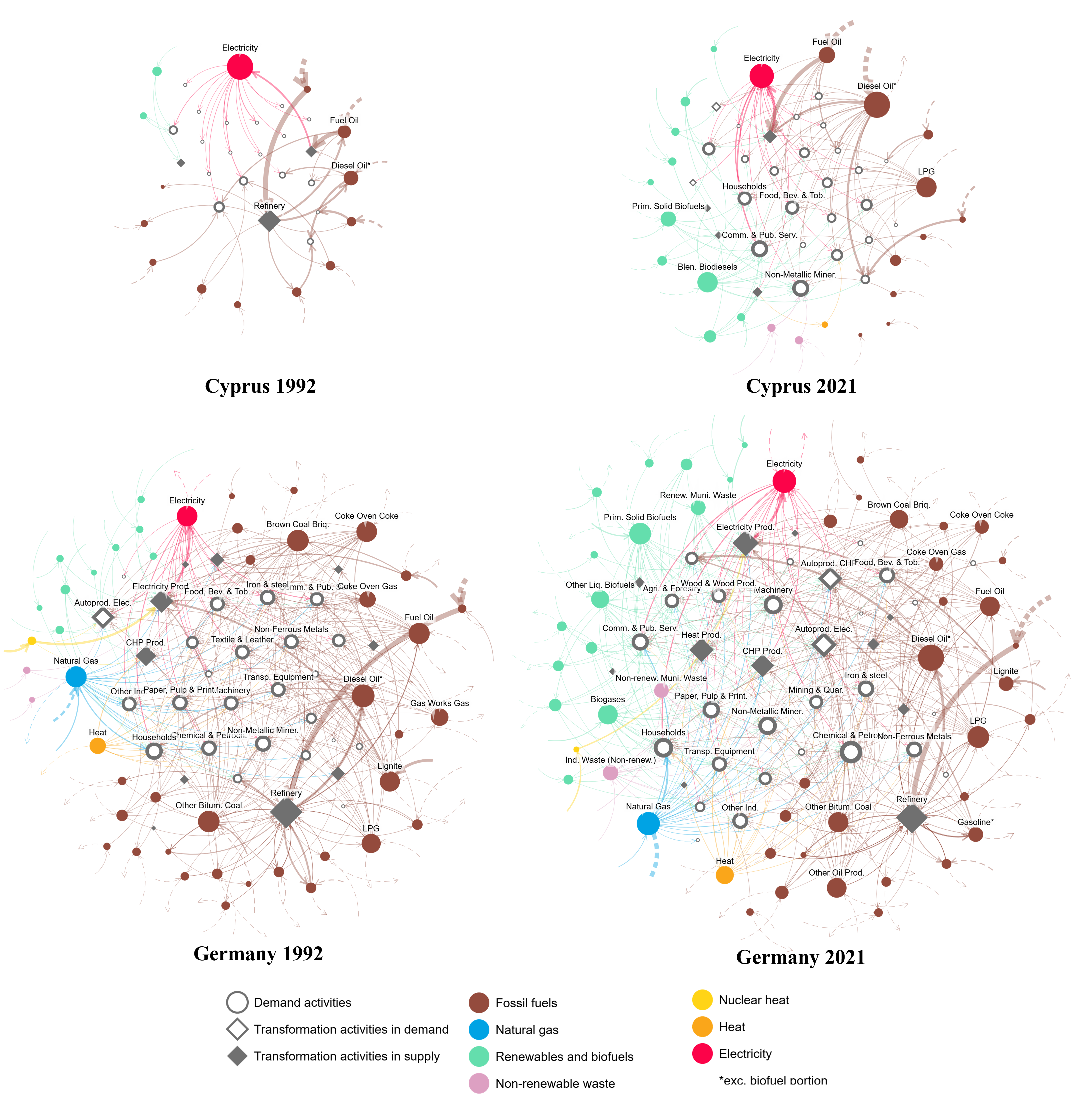}
    \caption{The evolution of the national energy system network of Cyprus and Germany from 1992 to 2021. The size of the node is proportional to degree centrality.}
    \label{fig:network_evolution}
\end{figure}

Figure \ref{fig:network_evolution} displays two extreme cases of the energy network evolution from 1992 to 2021: Cyprus and Germany. Node size reflects the importance of each node in the whole system in terms of connections (see Methods for more details).
Compared to Cyprus, Germany's network appears to be bigger, showcasing a higher number of nodes and edges, which reflects its historically diversified energy mix and advanced infrastructure.

Regarding Cyprus, in 1992, its energy system was predominantly dependent on fossil fuels, mainly oil derivatives (brown nodes), which powered electricity generation and a few final demand sectors. Electricity (red node), on the other hand, served as a central hub for most demand sectors, except transport. By 2021, the energy network diversified, with renewable sources like biofuels, wind, and solar (green nodes) emerging as alternative pathways for both electricity production and direct supply to final sectors. This shift increased the centrality of final demand sectors, indicating their growing reliance on diverse energy carriers. Nevertheless, fossil fuels still dominate the system, reflecting the challenges of transitioning for small, isolated economies.

In contrast, Germany had a more diversified energy mix in 1992, with coal, natural gas, and liquid fossil fuels playing significant roles. By 2021, renewables had expanded significantly in their contribution to both electricity generation and final demand. Despite the gradual phase-out of coal (and nuclear energy), fossil fuels remain important while renewables gradually gain importance.

\section*{Energy flows and the complexity of the European energy networks}

In the context of the energy transition, an initial growth of the system is expected due to the introduction of renewable energy sources, and potentially new activities that supply, transform and use these sources. This also requires the system to expand in terms of interactions. However, the phaseout of fossil fuels in favour of renewable clean energy towards the end of the transition could lead to an overall complexification or  simplification of the system. 
Furthermore, this introduction and phaseout do not take place discretely, but rather energy carriers and activities continuously vary the amount and composition of energy they use. Therefore, it is crucial to take into account not only changes in the (binary) links, but also the (weighted) flows passing through them and the effects of such changes throughout the system. 
Our goal here is to capture the complexification and de-complexification these processes give rise to, and the effects of complexity on the progress of the energy transition.

\subsection*{Energy flow patterns}

In this perspective, we aim to characterize European energy networks by examining recurrent and statistically significant patterns of energy flows. For this purpose, we model energy flows through the framework of random walkers\cite{picciolo2022weighted}. Each walker, essentially an indivisible unit of energy, traverses the network along its edges, choosing which outgoing link to take from a particular node with probability proportional to the associated link weight, i.e. the yearly aggregated energy flow. Walkers may also terminate their trajectories at nodes with more energy inflow than outflow, with probability proportional to this difference. This is represented schematically in Figure \ref{fig:walker_paths_representation} in Methods.

Thus, the energy flow
%Each walker represents an indivisible unit of energy that follows a unique path, beginning at a specific node and traversing the system without any loss of energy content. The walker leaves the system only at the final node of its trajectory, capturing how the amount of energy it represents exits the system at this point. In this framework, energy transfer
between nodes is captured by the number of walkers taking a step between them along their paths. This is essentially energy that is redistributed and therefore stays within the system.
%A fraction of these walkers may terminate their trajectory at the node they reach, representing an
The amount of energy leaving the system from a given node is represented by how many walkers terminate their paths there, and the amount of energy introduced to the system through a certain node is given by how many walkers start their path at it. %Meanwhile other walkers continue their journey through the network, representing the redistribution of energy across successive nodes until they eventually reach the end of their own paths (see Methods for theoretical details). 
Note that walkers are not specific to energy held in a particular physical form, as they can take paths that traverse several carrier nodes before leaving the system.

The advantage of the walker perspective over that of the data in its original form is that it avoids the double counting of energy flowing through the system. 
To illustrate this, consider imported electricity that is partially used for local electricity production, in turn producing more of the same carrier to be used by the same activity. In the yearly aggregated data, part of the originally introduced energy retraces a step in its path (from the carrier to the activity) despite no new energy having been introduced. From the perspective of the walkers, on the other hand, the amount of energy introduced is the amount of walkers that start their paths at the electricity carrier node, which can be decomposed into a the fraction that ends their path the first time they reach the electricity production activity, those that end their path the second time, and so on. Additionally, these fractions can be used to reconstruct the yearly aggregated flows, meaning that this information is not lost.

\subsection*{Effective transformation and the energy transition }

To measure the overall effect of the energy system from start to finish of the paths that walkers carry out, we introduce the concept of \textit{effective transformation}. It captures how many walkers introduced in the form of one type of energy carrier $\kappa$ leave the network in another type $\kappa'$, possibly the same as $\kappa$, regardless of the path through the energy system of a country from one to the other. This relates the forms in which energy is actually used (the final forms) to the forms of energy used indirectly (the introduced forms), all along the supply chain.

We can also further restrict paths to those where energy leaves from a particular activity node, describing how initial types are transformed into final ones used by that activity. This is the activity-level effective transformation. If we focus on energy leaving from any node in a specific group, such as final demand activity nodes, then we obtain the group-level effective transformation. %If we consider all paths within a country, regardless of which node walkers leave from (but not which form they leave in), the effective transformation is viewed at the country level.

At any of these levels of resolution, it is useful to visualise the effective transformation as a matrix where rows represent initial types in which energy enters the system, and the columns represent energy exiting the system in those same forms. Each element in the matrix is then the proportion of walkers that start and end in the carrier type corresponding to the row and column respectively.

Furthermore, the effective transformation can be used to obtain indicators for progress of the energy transition. For example, the share of non-renewable energy introduced to a country corresponds to the sum of elements in the non-renewable row of the country-level transformation matrix. 
The proportion of energy used in the form of a specific type is given by the sum of elements in the corresponding column. Note that these are alternative indicators whose relation depends on the structure of the effective transformation, and therefore of the energy system, in a way that is similar to how responsibility for emissions can be accounted for according to production or consumption. The same concepts apply to the group or activity level effective transformations (see eqs.\ref{eq:input} and \ref{eq:use} in Methods).

%\Cref{fig:type_effective_transformation_NO,fig:type_effective_transformation_IS} illustrate effective transformations in Norway and Iceland for 1991 and 2019 (later years are avoided due to the COVID-19 pandemic). Norway’s energy system remains largely unchanged, while Iceland shows clear progress in phasing out non-renewables. These figures also depict the total energy introduced (left) and used (top) for each group within the effective transformations.
%
%\begin{figure}[ht!]
%    \centering
%    \makebox[\textwidth][c]{
%    \includegraphics[width=0.4\linewidth]{NO_1992_eff_trans_type.png}
%    \includegraphics[width=0.4\linewidth]{NO_2019_eff_trans_type.png}
%    }
%    \caption{Effective transformation from input carrier types to final carrier types in Norway in 1992 (left) and 2019 (right). The total energy flows in the country are given at the top of each subfigure, and the curves on the left and top represent sums of rows (total energy introduced in a certain form) and of columns (total used in a certain form).}    
%    \label{fig:type_effective_transformation_NO}
%\end{figure}
%\begin{figure}[ht!]
%    \centering
%    \makebox[\textwidth][c]{
%    \includegraphics[width=0.4\linewidth]{IS_1992_eff_trans_type.png}
%    \includegraphics[width=0.4\linewidth]{IS_2019_eff_trans_type.png}
%    }
%    \caption{Effective transformation from input to final carrier types in Iceland in 1992 (left) and 2019 (right).}    
%    \label{fig:type_effective_transformation_IS}
%\end{figure}

\begin{figure}[ht!]
    \centering
    \makebox[\textwidth][c]{
    \includegraphics[width=0.4\linewidth]{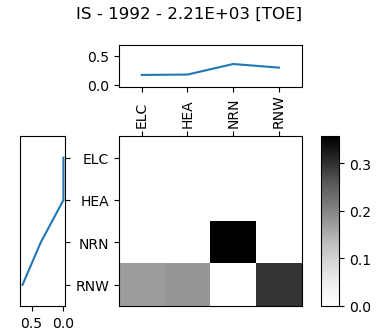}
    \includegraphics[width=0.4\linewidth]{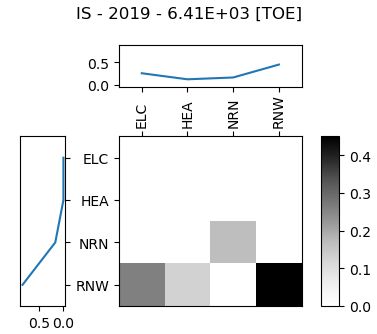}
    }
    \caption{Effective transformation from input to final carrier types in Iceland in 1992 (left) and 2019 (right). The total energy flows in the country are given at the top of each subfigure, and the curves on the left and top represent sums of rows (total energy introduced in a certain form) and of columns (total used in a certain form).}    
    \label{fig:type_effective_transformation_IS}
\end{figure}

For example, \cref{fig:type_effective_transformation_IS} illustrates the effective transformation in Iceland for 1991 and 2019 (later years are avoided due to the COVID-19 pandemic). These figures also depict the total energy introduced (left) and used (top) for each group within the effective transformations. Iceland shows clear progress in phasing out non-renewables, as seen by the reduction of both the introduced and used proportions of non-renewables.

\begin{figure}[ht!]
    \centering
    \includegraphics[width=0.9\linewidth]{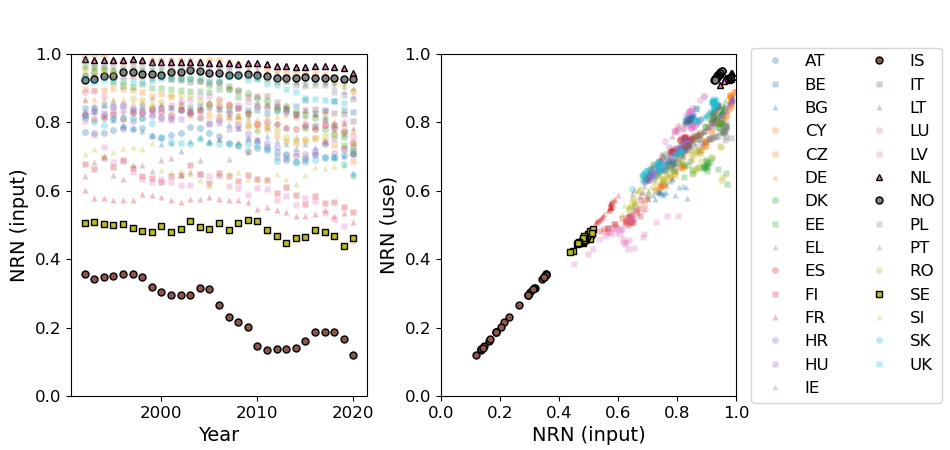}
    \caption{Proportion of non-renewable energy flows introduced to each country relative to total energy as a function of time, on the left, and the amount of energy used directly in the form of non-renewables compared to the amount introduced, on the right. Norway and Netherlands are highlighted as countries with the largest proportions of non-renewables, while Sweden and Iceland are those with the smallest proportions.}
    \label{fig:transition_progress}
\end{figure}

Using the JRC-ENERNET data (29 countries over 29 years), we calculate the effective transformation for each activity, groups of activities, and the entire country. Figure \ref{fig:transition_progress} (left) reports the share of non-renewable energy flows - relative to the total energy - introduced at the country level over time. Hence, a decreasing trend in these proportions can offer a first indication of the country’s progress in the energy transition. The right panel of Figure \ref{fig:transition_progress} compares the energy actually used in the form of non-renewables  with the energy initially introduced into the system in the same form. 
Direct use of non-renewables is slightly lower than the energy introduced, as indicated by points lying below the 45-degree line; this shows that measuring non-renewable use based solely on direct use leads to underestimate the non renewables that are actually required for the system to function. 

At a country level, most countries still need to decarbonise at least 50\% of their energy systems. Norway and the Netherlands are near the starting point of the energy transition (with over 90\% of energy flows from non-renewables), while Iceland and Sweden appear closer to its completion (with roughly 20\% and 50\% of energy flows from non-renewables, respectively). 
Due to their significant role as exporters, we could guess that much of Norway’s and the Netherlands’ non-renewable energy is not consumed domestically. However, this cannot be seen from \cref{fig:transition_progress} alone, but requires studying group-level effective transformations. 

\begin{figure}[ht!]
    \centering
   \includegraphics[width=0.9\linewidth]{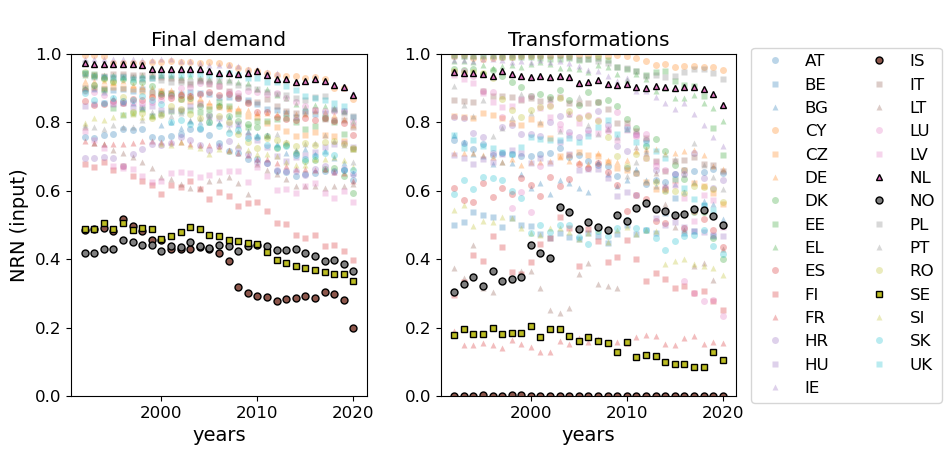}
    \caption{Progress of the energy transition for the final demand group, on the left, and for transformation activities, on the right, as a function of time for different countries.}
   \label{fig:transition_progress_types}
\end{figure}

For example, the row sums of the group-level effective transformation restricted to final demand yield the total amount of energy introduced in the form of each type of energy and used by final demand activities. In particular, focusing on the row corresponding to non-renewables reveals the reliance of final demand on non-renewables. This way, the left panel in \cref{fig:transition_progress_types} displays the contributions from non-renewables to the final demand activity group of different countries, while the right panel focuses on contributions to transformation activities. We observe that Norway’s final demand is comparable to highly decarbonised countries, 
whereas the Netherlands remains heavily carbon-intensive. Notably, transformation activities in Norway continue to increase dependence on introduced non-renewables. %Thus, the effective transformation at different levels of aggregation captures different perspectives on what makes a country advanced or delayed in its progress of the energy transition. However, the goal here is to show some common properties of how these perspectives interact with the complexity of the system rather than to choose a single criterion and argue that it is somehow better than others.

\subsection*{Heterogeneity of energy flow patterns}

Whereas the previous section examined how different energy paths contribute to the overall system flows, in this section we turn our attention to their heterogeneity. An energy system is considered more heterogeneous when the shares of paths within it are distributed more evenly. For example, if renewables are introduced, non-renewables lose part of their share to renewables, thus increasing the heterogeneity of the system. If non-renewables are subsequently phased out, then shares become concentrated around renewables, and the system becomes more homogeneous.

While many measures exist describing how evenly distributed shares of a total are, the Herfindhal-Hirschman Index (HHI) is often used in Economics to quantify market concentration \cite{rhoades1993herfindahl}. It ranges from $1/N$ in case of $N$ equal shares (i.e., perfect competition) to $1$, where all shares of the market are concentrated in one firm (i.e., monopoly). The HHI has also been applied to evaluate the diversification of other systems, such as financial portfolios and biological networks. It can be understood in a similar way to biodiversity, which is also often used to capture the complexity of ecosystems across or within trophic levels. Thus, to measure the heterogeneity of energy systems, we calculate the HHI using the shares of energy associated to the different paths captured in the effective transformation. Since the index takes its maximum value at lowest heterogeneity, we will refer to $1-\text{HHI}$ (see Methods for details).

\begin{figure}[h!]
    \centering
    \includegraphics[width=0.9\linewidth]{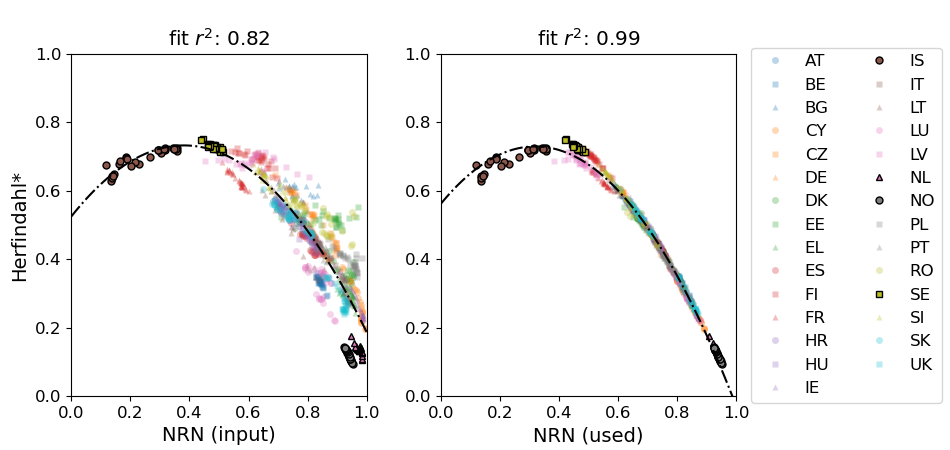}
    \caption{Herfindahl index calculated from proportions of paths in the type-level effective transformation, as a function of the introduced amount of non-renewables and the used amount of non-renewables on left and right respectively. All countries follow a parabolic trend with a higher $r^2$ when the control variable is the used amount of non-renewables than the introduced amount.}
    \label{fig:type_effective_transformation_herf}
\end{figure}

In Figure \ref{fig:type_effective_transformation_herf} we show the Herfindahl index measured on the effective transformation as a function of the proportion of non-renewables introduced (left) and used (right) in the system. Two main results emerge from the analysis of the energy transition. First, the evolution of heterogeneity follows a consistent parabolic trajectory across countries: it increases during the initial decrease of non-renewable sources at the start of the energy transition, when countries rely mostly on this type of energy carrier. This increase continues until close to $50\%$ of non-renewables, where the increase in complexity stagnates. No country except for Iceland has crossed this maximum complexity threshold, and following it Iceland shows a subsequent decrease in complexity as non-renewables continue to be phased out. Although Iceland may represent a peculiar case due to its relatively limited economic complexity, with a remarkably small and constant size of the energy network (between 50 and 60 nodes), the fact that all countries follow the same parabolic trajectory suggests that the peak of complexity may be reached with slightly different shares of NRN, but remains inevitable.

This trajectory is particularly consistent when the transition is evaluated in terms of energy used directly rather than the forms in which it is input upstream in the energy system, which may be attributed to the demand-driven characteristics of European economies. Second, the reduction in heterogeneity observed at the end of the transition is markedly smaller than the initial increase. This outcome could be interpreted in terms of a net rise in systemic complexity once electricity and renewables are jointly incorporated to replace non-renewables. 

\begin{figure}[h!]
    \centering
    \includegraphics[width=0.9\linewidth]{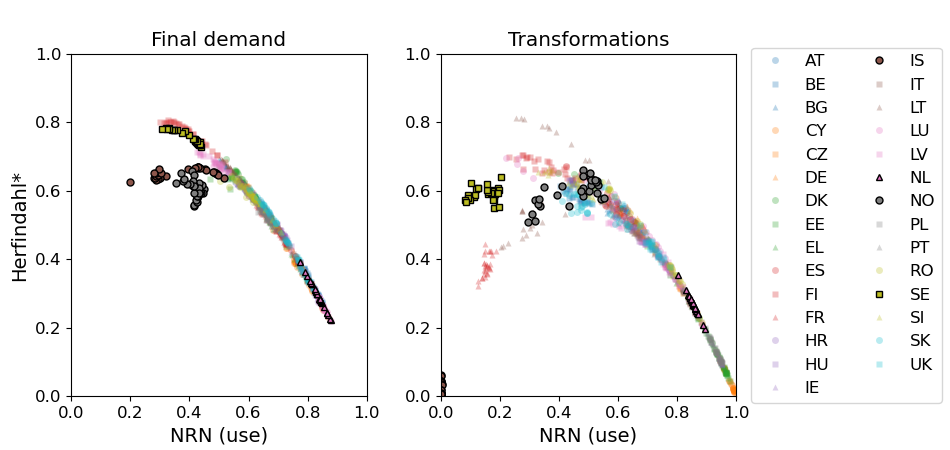}
    \caption{Herfindahl index calculated for the effective transformation at the level of activity groups, for final demand on the left and transformation activities on the right. Both are plotted as a function of the amount of non-renewables used by the group.}
    \label{fig:activity_group_herf}
\end{figure}

At the level of activity groups (Figure \ref{fig:activity_group_herf}), a general parabolic trend is also observed, although it is much less consistent among  countries, suggesting that the same activity in different countries can require a higher or lower level of complexity to  move away from fossil fuels, probably because of structural differences in the economic fabric. 
Below approximately 50\% of non-renewables, the trajectories become more heterogeneous across groups; nevertheless, the general pattern persists, with heterogeneity first increasing and then declining once a threshold proportion of non-renewables is reached. The parabolic structure of the Herfindahl index remains discernible even when measuring the Herfindahl index at a more granular level, in terms of all carriers instead of carrier types, as can be seen in Figure \ref{fig:node_level_effective_transformation_herfindahl_NRNRNW} in Methods.

\begin{figure}[ht!]
    \centering
    \includegraphics[width=0.9\linewidth]{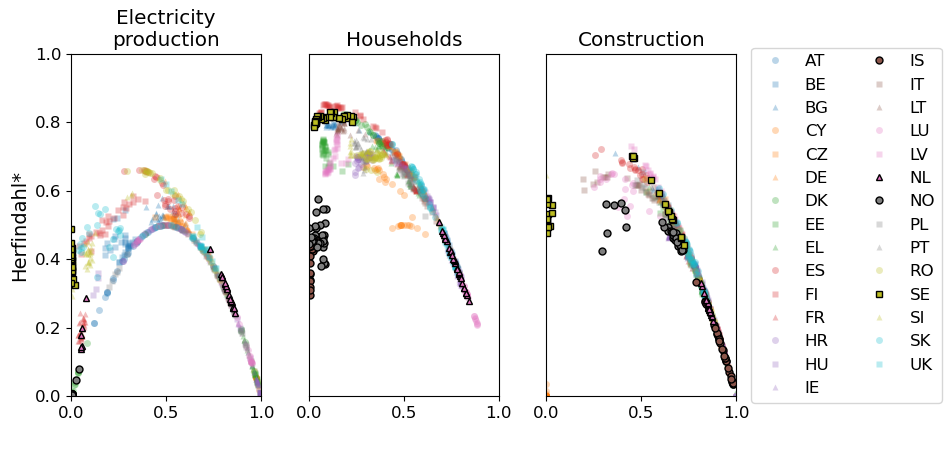}
    \caption{Herfindahl index measured from shares of carrier types used by three different activities, as a function of the proportion of non-renewables used by each.}
    \label{fig:herfindahl_by_activity}
\end{figure}

Finally, Figure \ref{fig:herfindahl_by_activity}  reports the Herfindahl index computed with respect to the proportions of carrier types in the energy use of selected activities. That is, they are caclulated from activity-level effective transformations. We focus on three representative cases - electricity production, households, and construction - while the complete set of activities is shown in supplementary figure 3 (\cref{supp-fig:activity_level_herfindahl}). Across countries, the index values systematically follow two parabolic trajectories, indicating that changes in complexity remain confined within a specific range. The emergence of these two distinct parabolic patterns can be explained by differences in supply-chain flexibility: some activities can be more easily adapted to alternative carriers with minimal modifications, whereas others require substantial modifications, resulting in higher levels of complexity for the system, which is maintained throughout the transition.

\subsection*{Energy transition patterns}

\begin{figure}[ht!]
    \centering
    \includegraphics[width=0.9\linewidth]{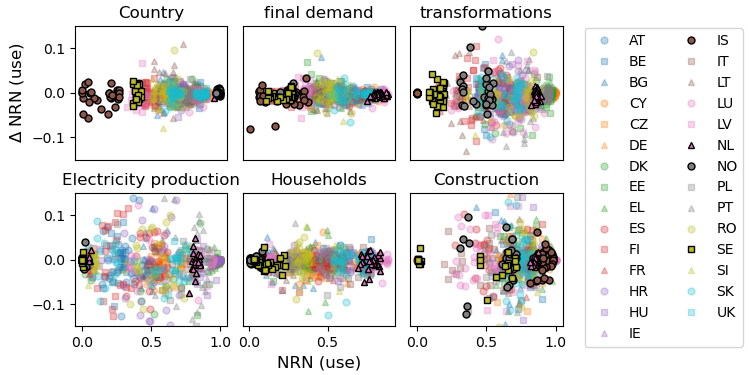}
    \caption{Average speed of the transition as a function of its progress, the average proportion of non-renewables involved in inputs, use, and specific to construction, households and electricity production activities. }
    \label{fig:speed_vs_progress}
\end{figure}

In the previous section we have studied how the transition affects the heterogeneity of paths that energy can take in the system of each country, induced by the presence of new energy forms. We now focus on the effects that this heterogeneity has on the progress of the energy transition.
In Figure \ref{fig:speed_vs_progress} we show the yearly variation
 in the proportion of used non-renewables, indicating how fast the transition advances or retreats in a year, with respect to the proportion of used non-renewables. 
On the top of Figure \ref{fig:speed_vs_progress} we report results at country level and for two activity groups (the same of \cref{fig:activity_group_herf}), while on the bottom we focus on activities (the same of \cref{fig:herfindahl_by_activity}).

We observe that the speed of the transition is largely independent of the progress of the transition. We have measured the average speed at which non-renewables are removed is at most $2\%$ per year, meaning that the transition would take about $50$ years, far beyond current goals. Instead, fluctuations dominate the behaviour of the system rather than the average speed. Additionally, the construction and electricity production activities actually decarbonise more slowly as the proportion of non-renewables decreases, and both present notably larger fluctuations than the others.

\begin{figure}[ht!]
    \centering
    \includegraphics[width=0.9\linewidth]{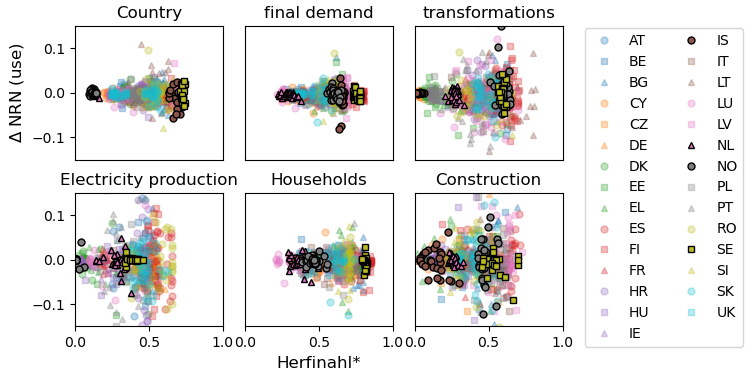}
    \caption{Average speed of the transition, measured by the change in the proportion of non-renewables, as a function of the Herfindahl index of the effective transformation, use table, and specific to Construction, Households and Electricity production activities. }
    \label{fig:speed_vs_herfindahl}
\end{figure}

To better characterise the fluctuations, we hypothesize that in a more heterogeneous system there is more room for rapid changes, so we would expect the Herfindahl index to influence the size of fluctuations of the speed at which the transition takes place. In \cref{fig:speed_vs_herfindahl} we show the change in the same proportions of non-renewables shown in \cref{fig:speed_vs_progress}, now as a function of the Herfindahl index instead of the proportion of non-renewables.

In this case, the picture is much clearer. As the heterogeneity measured by the Herfindahl index increases, the size of fluctuations increases, as expected. Fluctuations are not necessarily larger, but their distribution becomes much wider. Thus, transitions are expected to cross a ``turbulent'' zone close to where the Heterogeneity of the system is highest, which will be characterised by large changes with relatively slow progress. 

\section*{Conclusions}

In this paper, we have applied techniques from network science to study shifts in energy systems of European countries in the last $30$ years, using data from the new JRC-ENERNET database. Using random walkers on networks, the roles of activities in the introduction, transformation, and use of different forms of energy were decomposed into different paths that energy takes through the system. With these paths, we first constructed what we call effective transformations, representing the proportion of energy that is introduced to the system in an initial form and, after an arbitrary amount of transformations, used in a final one, thereby leaving the system.
With the proportions of different energy paths, we proceeded to calculate their heterogeneity according to the Herfindahl index. The proportions of energy were interpreted as shares, obtaining a measure of heterogeneity that was applied to the level of countries, groups of activities, and individual activities to identify patterns and variety.
In other words, while initially the transition drives an increase in complexity, after a certain and remarkably ubiquitous threshold of share in non-renewables, complexity starts decreasing. It is worth noting that, while at the activity level, countries may exhibit different parabolic trajectories, some ending at the same level of complexity, others requiring a higher level to progress in the transition, when considered at the aggregate level they appear to follow the same trajectory.  This remarkably common behaviour reflects the need for auxiliary energy forms beyond only non-renewables, such as electricity or heat, to fully carry out the transition. 

Moreover, we found that the heterogeneity influences the size of fluctuations in the proportion of non-renewable energy, thus influencing further progress and pace of the transition. More heterogeneous (and complex) energy networks are found to exhibit larger fluctuations (both positive and negative) in their proportion of non-renewables, making them more apt to replace non-renewables with renewables but also more susceptible to reverse this trajectory. 
Accelerating the transition should therefore be pursued through a policy aimed at promoting greater diversification and integration across the whole gamut of vectors and activities, rather than a monosectoral policy aimed at replacing, concentrating and scaling up vertically.

In future work, we aim to continue to develop and refine the relation between heterogeneity and shifts in energy systems. In particular, the model to reconstruct the behaviour of individual activities and energy forms in the system is based on commonly used but also somewhat simplistic assumptions. Whether it is worthwhile to improve on these assumptions is yet to be seen. Additionally, the common behaviour of different countries, with few outliers begs the question of whether it is possible and how to change this behaviour. For example, as we find that heterogeneity increases the size of fluctuations in the proportion of non-renewables, it may be possible to accelerate the transition by regulating heterogeneity.

\section*{Methods}
\label{sec:methods}

\subsection*{Data}

Abbreviations for country names are: Austria (AT),
Belgium (BE),
Bulgaria (BG), 
Cyprus (CY),
Czechia (CZ),
Germany (DE),
Denmark (DK),
Estonia (EE),
Greece (EL),
Spain (ES),
Finland (FI),
France (FR),
Croatia (HR),
Hungary (HU),
Ireland (IE),
Iceland (IS),
Italy (IT),
Lithuania (LT), 
Luxembourg (LU), 
Latvia (LV),
Netherlands (NL), 
Norway (NO), 
Poland (PL), 
Portugal (PT),
Romania (RO), 
Sweden (SE), 
Slovenia (SI), 
Slovakia (SK), 
United Kingdom (UK).
%given in \cref{tab:country_codes}. The node classification is given in \cref{tab:node_types}
%\begin{table}[!ht]
%\centering
%\small
%\begin{tabular}{l | c } 
%\textbf{Name} & \textbf{Acronym} \\ [1ex] 
%\hline \hline  
%Austria     & AT \\
%Belgium     & BE\\
%Bulgaria    & BG\\ Cyprus      & CY\\
%Czechia     & CZ\\
%Germany     & DE\\ Denmark     & DK\\
%Estonia     & EE\\ Greece      & EL\\ Spain       & ES\\
%Finland     & FI\\
%France      & FR\\ Croatia     & HR\\ Hungary     & HU\\ Ireland     & IE\\ Iceland     & IS\\ Italy       & IT\\ Lithuania   & LT\\ Luxembourg  & LU\\ Latvia      & LV\\ Netherlands & NL\\ Norway      & NO\\ Poland      & PL\\ Portugal    & PT\\
%Romania     & RO\\ Sweden      & SE\\ Slovenia    & SI\\ Slovakia    & SK\\ United Kingdom  & UK\\
% \hline
%\end{tabular}
%  \caption{Country names and acronyms.}
%  \label{tab:country_codes}
%\end{table}
%\include{classification_table}

\begin{table}[ht!]
\centering
\small
\begin{tabular}{l|l|l|p{0.7\textwidth}}
\textbf{Actor} & \textbf{Type} & \textbf{Flow} & \textbf{Node} \\
\hline \hline
carrier & ELC & engine & Electricity \\
carrier & HEA & engine & Heat, Nuclear heat \\
carrier & NRN & engine & Additives and oxygenates (exc. biofuel portion), Anthracite, Aviation gasoline, Bitumen, Blast furnace gas, Brown coal briquettes, Coal tar, Coke oven coke, Coke oven gas, Coking coal, Crude oil, Ethane, Fuel oil, Gas coke, Gas oil and diesel oil (exc. biofuel portion), Gas works gas, Gasoline-type jet fuel, Industrial waste (non-renewable), Kerosene-type jet fuel (exc. biofuel portion), Lignite, Liquefied petroleum gases, Lubricants, Motor gasoline (exc. biofuel portion), Naphtha, Natural gas, Natural gas liquids, Non-renewable municipal waste, Oil shale and oil sands, Other bituminous coal, Other hydrocarbons, Other kerosene, Other oil products, Other recovered gases, Paraffin waxes, Patent fuel, Peat, Peat products, Petroleum coke, Refinery feedstocks, Refinery gas, Sub-bituminous coal, White spirit \\
carrier & RNW & engine & Ambient heat (heat pumps), Biogases, Blended biodiesels, Blended biogasoline, Charcoal, Geothermal, Hydro, Other liquid biofuels, Primary solid biofuels, Pure bio jet kerosene, Pure biodiesels, Pure biogasoline, Renewable municipal waste, Solar photovoltaic, Solar thermal, Tide, wave, ocean, Wind \\
activity & final demand & sinks & Agriculture \& forestry, Coal liquefaction plants, Commercial \& public services, Construction, Domestic aviation, Domestic navigation, Fishing, Food, beverages \& tobacco, Households, International aviation, International maritime bunkers, Iron \& steel, Machinery, Mining \& quarrying, Non-ferrous metals, Non-metallic minerals, Other industry, Other sectors, Other transport, Paper, pulp \& printing, Pipeline transport, Rail, Road, Textile \& leather, Transport equipment, Wood \& wood products \\
activity & transformations & engine & Autoproducer CHP , Autoproducer Electricity, Autoproducer Heat, BKB \& PB plants, Blast furnaces, Blended natural gas, CHP production, Charcoal production plants, Chemical \& petrochemical, Coke ovens, Electric boilers, Electricity production, Gas works, Gas-to-liquids plants, Heat production, Heat pumps, Liquid biofuels blended, Patent fuel plants, Pumped hydro, Refinery \\
activity & primary production & sources & PPRD + carrier node \\
activity & imports & sources & IMP + carrier node \\
activity & exports & sinks & EXP + carrier node \\
\hline
\end{tabular}
\caption{Node classification.}
\label{tab:node_types}
\end{table}

\subsection*{Energy introduction}

The data contained in JRC-ENERNET consists of total yearly flows of energy in different countries, establishing flows from a given activity in the form of each carrier $W_{ac}$ (from an activity node $a$ to a carrier node $c$) and to a given activity in the form of each carrier $W_{ca}$ (flows from a carrier node $c$ to an activity node $a$). It is useful to interpret $W_{ac}$ and $W_{ca}$ as elements of the off-diagonal blocks of a matrix $W_{ij}$ where $i$ and $j$ can be either activities or carriers, and elements in the diagonal blocks $W_{aa'}$ and $W_{cc'}$ are always $0$.

The total energy arriving at a given activity over the course of a year from within the energy system of a particular country is then $S^{in}_a = \sum_c W_{ca} = \sum_j W_{ja}$, and the total leaving it that stays within the system, flowing to other nodes, is $S^{out}_a = \sum_c W_{ac} = \sum_j W_{aj}$. For transformation activities, we expect that $S^{in}_a \geq S^{out}_a$ as the missing energy $S^{in}_a - S^{out}_a$ leaves the energy system during transformation processes, for example in the form of dissipation losses. Because of the structural properties of the network, shown in \cref{fig:network_representation}, all the energy that arrives at sink activities $S^{in}_a$ leaves the energy system because there are no pathways for energy to continue from them $S^{out}_a = 0$. Source activities, on the other hand, only present outward flows, which must then come from outside the system.

This idea can be used as a principled way to find where and how much energy enters or leaves the system based on the flows in the data. For any node $i$ in the network, whether activity or carrier, if the difference $S^{in}_i - S^{out}_i$ is positive then it represents the amount of energy $L_i$ that leaves the system from node $i$. If the difference is negative, on the other hand, then it is the amount of energy entering the system at the node $F_i$.
\begin{equation}
    L_i = \max(S^{in}_i - S^{out}_i, 0) ~~~~~~~~ F_i = \max(S^{out}_i - S^{in}_i, 0) \, .
    \label{eq:leaving_or_entering}
\end{equation}
Therefore, the total amount of energy arriving at a node $i$ from both within and outside the system is $F_i + S^{in}_i$. 

\subsection*{Redistribution and use}

Out of the total energy that a node receives $F_i + S^{in}_i$, $L_i$ is lost and $W_{ij}$ continues within the system to another node $j$, defining the proportions $q_i$ and $P_{ij}$ respectively,
\begin{equation}
    q_i = \frac{L_i}{F_i + S^{in}_i} ~~~~~~~~ P_{ij} = \frac{W_{ij}}{F_i + S^{in}_i} \, .
    \label{eq:sink_and_redist_prob}
\end{equation}
Note that these proportions are normalised together, $q_i + \sum_j P_{ij} = 1$, for each node $i$. This means that all energy arriving at each node over the course of a year is either lost or redistributed to others according to the proportions \cref{eq:sink_and_redist_prob}.

We will now assume that the proportions of energy lost and redistributed apply not only over the course of an entire year, but also for energy received at any particular moment in the year. That is, if node $i$ receives a flow $K_i$, regardless of whether it comes from within or outside the system, or the carriers it comes from, then it will redistribute this energy to other nodes in amounts $K_i P_{ij}$, and $K_i q_i$ will be used in the process. This is inspired by the model for weighted network motifs of \cite{picciolo2022weighted}. 

The assumption that energy is always redistributed by a node in the same way over the course of a year in a given country gives a way to reconstruct the flows throughout the year. For this, consider an amount of energy $K_i$ introduced to a node $i$. Because the loss and redistribution probabilities \cref{eq:sink_and_redist_prob} are normalised, we can write
\begin{equation}
    K_i = K_i (q_ i + \sum_j P_{ij}) = K_i q_i + \sum_j K_i P_{ij} \, .
\end{equation}
The first term $K_i q_i$ is the amount of energy that is used at node $i$, while $K_i P_{ij}$ are proportions that arrive at nodes $j$ from node $i$. For each of these, we can again write
\begin{equation}
    \begin{aligned}
        &K_i P_{ij} = K_i P_{ij} q_j + \sum_k K_i P_{ij} P_{jk} \\
        \Rightarrow K_i  = K_i q_i + \sum_j &K_i P_{ij} = K_i q_i + \sum_j K_i P_{ij} q_j + \sum_j K_i {P^2}_{ij} \, .
    \end{aligned}
\end{equation}
We now recognise the amount of energy used out of what arrives at node $i$, $K_i q_i$, the amount used at each node $j$ from redistribution in one step $K_i P_{ij} q_j$, and the amount of energy that arrives to other nodes in two redistributions steps $K_i {P^2}_{ij}$ where ${P^2}_{ij} = \sum_k P_{ik} P_{kj}$ is the second matrix power of $P$. 

\begin{figure}
    \centering
    \includegraphics[width=1\linewidth]{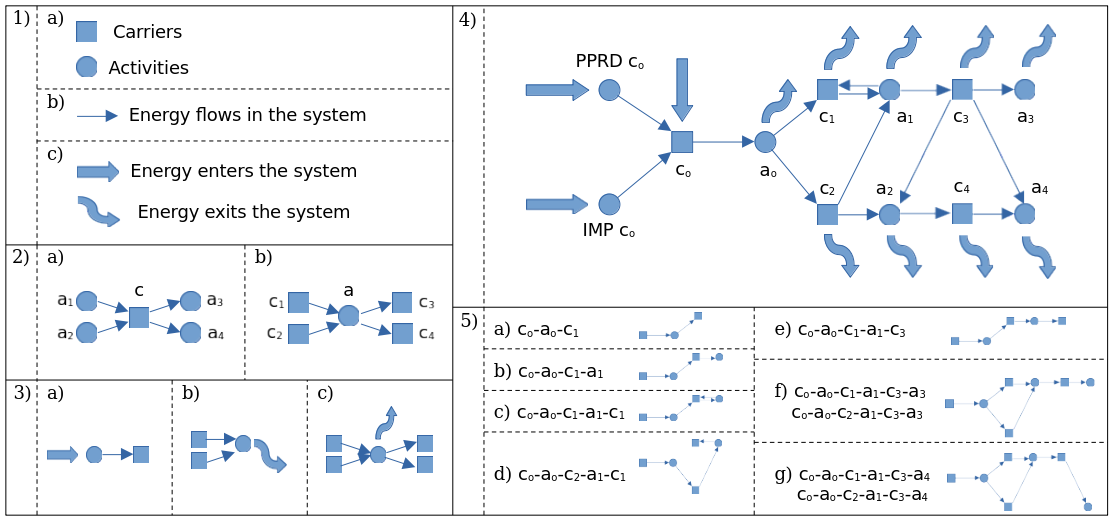}
    \caption{1a) Nodes in the network are either carriers or activities. 1b) Weighted links in the network represent flows of energy (in tons of oil equivalent) and occur only from carriers to activities or activities to carriers. 1c) Due to energy conservation at each node, energy enters the system at nodes where the total outflow to other nodes is larger that its total inflow from other nodes, and exits the system at nodes where the total outflow to other nodes is smaller than its total inflow from other nodes.\\
    2a) The physical form in which energy is found for flows from activities to a particular carrier and from that carrier to other activities is that carrier. For example, if $c$ is crude oil, then flows from $a_1$ and $a_2$ to node $c$ and flows to $a_3$ and $a_4$ are flows of crude oil. 2b) An activity can receive energy flows from different carriers (in different forms) and output energy in the form of other carriers. Activities can also output energy to a carrier they receive flows from. For example, if $a$ is a refinery, it might receive energy flows from crude oil and natural gas and produce gasoline and jet fuel. If $a$ is electricity production, it might receive and output energy to the electricity carrier node, among others.\\
    3a) If energy enters the system through an activity, it must be either imports or primary production (also known as source activities) of a particular carrier. This energy enters the system in the form of that carrier and therefore the only flows from a source activity to the carrier node. Source activities cannot receive energy from other nodes in the system. Energy can also enter the system through any carrier node. 3b) If energy exits the system through an activity which has no output to other nodes, this must be a final demand or export activity (also known as sink activities). Exports, like source activities, are of a specific carrier and therefore receive flows from only one carrier node. 3c) An activity node that has outputs to other nodes and from which energy exits the system must be a transformation activity. If energy exits the system through a carrier node, it must receive energy from other activities, and can have outgoing flows.\\
    4) An example network with energy entering only as carrier $c_o$ (from both source activities and the carrier node $c_o$). Energy exits the system from all other nodes. The activity $a_3$ can be an export activity (export of $c_3$) or a final demand activity, but $a_4$ must be a final demand activity as it receives energy from two different carriers. All other activities must be transformation activities as they have both incoming and outgoing flows to other nodes.\\
    5) Examples of sequences of steps that contribute to different types of paths. Paths are classified according to the form in which they enter the system $c_i$, the form in which they exit it $c_f$ and the node from which energy exits $n$. This type of path is denoted $c_i \rightarrow c_f (n)$. Examples a), c) and d) are paths of type $c_o \rightarrow c_1 (c_1)$. b) is a path $c_o \rightarrow c_1 (a_1)$. e) is $c_o \rightarrow c_3(c_3)$, both examples from f) are $c_o \rightarrow c_3(a_3)$ and both examples from g) $c_o \rightarrow c_3(a_4)$.}
    \label{fig:walker_paths_representation}
\end{figure}

If we continue expanding higher order terms by introducing the used and redistributed energy fractions, we obtain
\begin{equation}
    K_i = \sum_{t=0}^{\infty} \sum_j K_i {P^t}_{ij} q_j = \sum_j K_i {(I - P)^{-1}}_{ij} q_j \, .
\end{equation}
This expression can be interpreted as requiring the total energy $K_i$ introduced at a node $i$ to be used by any node $j$ in any number of redistribution steps $t$ in amounts $K_i {P^t}_{ij} q_j$. Thus, over any amount of steps in total means that node $j$ contributes a use $K_i {(I - P)^{-1}}_{ij} q_j$ of the energy $K_i$ introduced to node $i$. A graphical representation of the possible paths that energy takes through the system giving rise to these fractions is shown in \cref{fig:walker_paths_representation}.

Because of the block structure of the network, one-step paths $P_{ij}$ will go to carriers, if $i$ is an activity, or to activities if $i$ is a carrier. Two step paths ${P^2}_{ij}$ will return to the same class of node as $i$, and in general this will be the case for any even number of steps ${P^{2t}}_{ij}$. Odd number of steps ${P^{2t+1}}_{ij}$, on the other hand, will lead to nodes of the other class. It is therefore useful to separate into even and odd numbers of steps
\begin{equation}
    \begin{aligned}
        K_i = \sum_{t=0}^{\infty} K_i \sum_j {P^t}_{ij} q_j = \sum_j \sum_{t=0}^{\infty} K_i ({P^{2t}}_{ij} + {P^{2t+1}}_{ij})q_j \\
        = \sum_j K_i {(I - P^2)^{-1}}_{ij}q_j + \sum_j K_i {[(I - P^2)^{-1} P]}_{ij}q_j .
    \end{aligned}
\end{equation}

This way, if $i$ is a carrier node $c$, then even numbers of steps reach other carrier nodes $c'$ and odd ones activities $a$,
\begin{equation}
    K_c = \sum_{c'} K_c {(I - P^2)^{-1}}_{cc'}q_{c'} + \sum_a K_c {[(I - P^2)^{-1} P]}_{ca}q_a \, .
    \label{eq:initial_carrier_use_final_carriers_and_activities}
\end{equation}
This represents how energy introduced in the form of a specific carrier $c$ is used up by the system downstream through the supply chain. Out of a total amount $K_c$ introduced, $K_c{(I-P^2)^{-1}}_{cc'}q_{c'}$ arrives at a carrier node $c'$ and is used up there, and $[(I-P^2)^{-1}P]_{ca}q_a$ arrives at the activity node $a$ and used there. This way, the total amount of energy introduced is accounted for through its use by carriers and activities together.

Moreover, the use by an activity $a$ can be seen as a sum of contributions of energy arriving at carriers $c'$, but instead of being used they are redistributed once to the activity node $a$ and used there
\begin{equation}
    K_c[(I-P^2)^{-1}P]_{ca}q_a = \sum_{c'} K_c {(I - P^2)^{-1}}_{cc'} P_{c'a} q_a
    \label{eq:initial_carrier_use_by_activity_as_final_carriers}
\end{equation}
Because the flows from carrier $c'$ to activity $a$ represent energy in the form of $c'$, \cref{eq:initial_carrier_use_by_activity_as_final_carriers} can be interpreted as the use of energy in the form of different carriers $c'$ by an activity $a$, where these carriers were originally introduced in the form of $c$.

If \cref{eq:initial_carrier_use_by_activity_as_final_carriers} is inserted in \cref{eq:initial_carrier_use_final_carriers_and_activities}, we obtain contributions from the different final forms $c'$ in which a particular initial form of energy $c$ is used
\begin{equation}
    K_c = \sum_{c'} \left\{  K_c {(I - P^2)^{-1}}_{cc'} q_{c'} + \sum_a K_c {(I - P^2)^{-1}}_{cc'} P_{c'a} q_a \right\} \, .
    \label{eq:initial_carrier_use_as_final_carrier}
\end{equation}
The use of the initial carrier $c$ in the form of each possible final carrier $c'$ then comes both from losses intrinsic to the carrier node $c'$, $K_c {(I - P^2)^{-1}}_{cc'} q_{c'}$, and to activities $a$ that use the final carrier directly, $ K_c {(I - P^2)^{-1}}_{cc'} P_{c'a} q_{a}$. 

This way, when flows are introduced to the system in the form of a specific carrier, we can decompose how it is used into contribution from how this energy form leaves the system, accounting for both the final forms in which it is used and the activities and losses that lead to it exiting the energy system of a particular country.

However, we are not free to choose how energy enters the system, which we have assumed to be from a carrier node in the previous reasoning. The amount of energy introduced to each node is given by the network data according to \cref{eq:leaving_or_entering}. In particular, we have noted that energy must enter through source activities, as these only exhibit flows into the system which must come from the environment or other countries. Nevertheless, source activities import or produce a single carrier each, meaning that in fact we do know the form in which energy is introduced to the system. Moreover, this energy is completely transferred to the carrier node of the form in which energy is produced or imported, and these activities do not play any further role in the use or transformation of energy. Therefore we can imagine that they are actually external to the energy system, and that the total externally introduced energy to the carrier node $c$ is $K_c = F_c + F_{IMP~c} + F_{PPRD~c}$.

Beyond energy from the imports and primary production activities $F_{IMP~c}$ and $F_{PPRD~c}$, and possibly directly to the respective carrier nodes $F_c$, energy can only potentially enter through transformation activities where $S^{out}_a > S^{in}_a$. This is because sink activities, namely final demand and exports, only receive energy meaning their outgoing flows can never be larger than incoming ones. For the transformation activities, we expect that $S^{in}_a > S^{out}_a$ most of the time due to use of energy in transformations, so energy should not be introduced through these nodes. If it were, we would not be able to identify the initial form in which it is introduced, only the final forms in which it is used and the activities that use it. Assuming \cref{eq:leaving_or_entering} yields only one transformation activity where this is the case, responsible for less than $1\%$ of the total energy introduced to the country over the course of a year each time it does. We therefore remove these flows, and have verified that our results are not sensitive to this choice.

With all flows introduced to carrier nodes, we can return to their use by different activities in different final forms according to \cref{eq:initial_carrier_use_by_activity_as_final_carriers}. The total flows $F = \sum_c K_c$ introduced to an energy system of a country over the course of a year then leave the system through losses associated to carriers themselves or use by activities according to
\begin{equation}
    F = \sum_c \sum_{c'} \left\{ K_c {(I - P^2)^{-1}}_{cc'} q_{c'} + \sum_a K_c{(I - P^2)^{-1}}_{cc'} P_{c'a} q_a \right\} \, .
    \label{eq:total_introduced_losses}
\end{equation}
The proportion of energy, out of all flows in the system, that is introduced in the form of a carrier $c$ and is lost in the form $c'$, denoted $Q_{cc'}$, and the proportion that is introduced in the form $c$ and is used by an activity $a$ in the form $c'$, symbolised with $P_{cc'a}$, are then
\begin{equation}
    Q_{cc'} = p_c {(I - P^2)^{-1}}_{cc'} q_{c'} ~~~~~~~~ P_{cc'a} = p_c {(I - P^2)^{-1}}_{cc'} P_{c'a} q_{c'} \, ,
\end{equation}
where $p_c = K_c / F$ is the proportion of energy introduced in the form of $c$, out of the total $F$ introduced. Note that these expressions result from dividing \cref{eq:total_introduced_losses} by $F$, guaranteeing that these proportions are normalised $\sum_{cc'} Q_{cc'} + \sum_{cc'a} P_{cc'a} = 1$. This essentially means that together they account for all possible paths that energy can take through the system.

As discussed before, the carriers introduced or removed play a role in the transition according to the energy flows (not just the presence or absence of the node), now captured by the proportions of paths. However, the carriers found in each country may vary drastically, for example due to the presence of different activities without any direct significance on the energy transition. 
We therefore aggregate the proportions of paths into the possible pairs of groups of carriers (i.e. RNW, NRN, HEA and ELC), which we will denote $\kappa,\kappa'$ to distinguish it from the pair of carriers $c,c'$,
\begin{equation}
    Q_{\kappa \kappa'} = \sum_{\substack{c \in \kappa \\ c' \in \kappa'}} Q_{cc'} ~~~~~~~~ P_{\kappa \kappa' a} = \sum_{\substack{c \in \kappa \\ c' \in \kappa'}} P_{cc'a} \, .
\end{equation}

\subsection*{Energy system characterisation}
\label{sec:sys_charac}

The effective transformation captures how much of each energy carrier is introduced into the system, and how much leaves it in each form. This is calculated as the total proportion of energy introduced into the system in the form $c$ and leaving it in the form $c'$, aggregated over loss from the carrier node $c'$ itself and use of the final carrier by activities $a$, regardless of how many intermediate transformations the carrier underwent from initial to final form
\begin{equation}
    T_{cc'} = Q_{cc'} + \sum_a P_{cc'a} \, .
    \label{eq:effective_transformation_appendix}
\end{equation}

\begin{figure}
   \centering
    \includegraphics[width=0.7\linewidth]{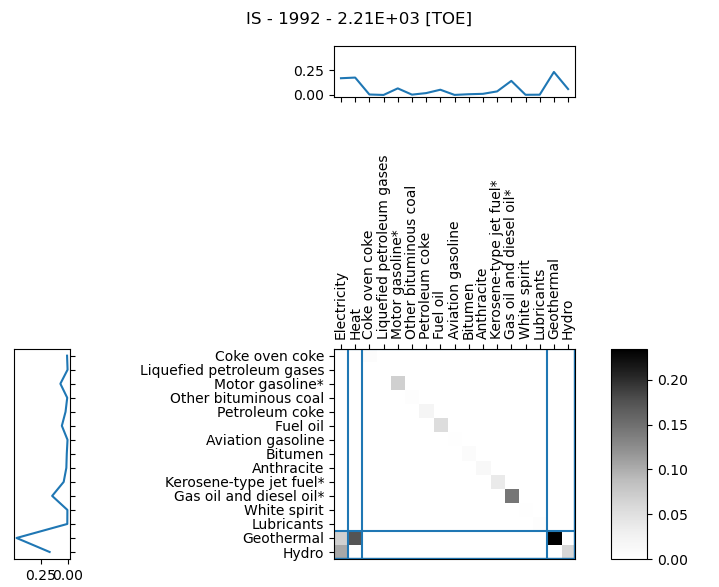}
    \caption{Effective transformation from input to final carriers in Iceland in 1992.}
    \label{fig:full_effective_transformation_IS_92}
\end{figure}

\begin{figure}
   \centering
    \includegraphics[width=0.7\linewidth]{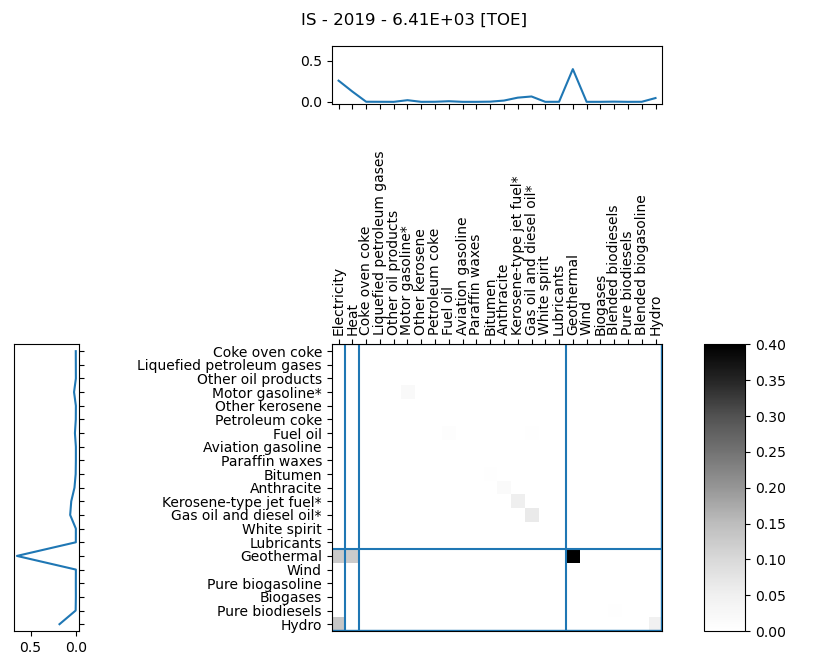}
    \caption{Effective transformation from input to final carriers in Iceland in 2019.}
    \label{fig:full_effective_transformation_IS_19}
\end{figure}

For example, the effective transformation and total amounts of introduced and used energy in Iceland in 1992 is shown in \cref{fig:full_effective_transformation_IS_92}. The shade of grey indicates the proportion of these flows in the entire system, constituting a total given in the title (in this case, $2.21 \times 10^3$ TOE). We see that the highest proportion of flows is of energy that enters and leaves the system in the form of 
geothermal and hydro power, used both in their own form and to produce heat and electricity. Additionally, many different non-renewables are used, but their proportion in the system is quite low compared to renewables. In 2019, shown in \cref{fig:full_effective_transformation_IS_19}, Iceland has managed to further lower the involvement of non-renewables and focus its dependence on Geothermal.


The effective transformation can also be understood as an average of effective transformations associated to each activity $a$, and direct loss. For this, note that the total energy used by an activity is $P_a = \sum_{cc'} P_{cc'a}$. Thus the proportion of this energy that is introduced in a form $c$ and leaves in a form $c'$, namely the effective transformation of the activity, is
\begin{equation}
    T^a_{cc'} = \frac{P_{cc'a}}{P_a} \, .
\end{equation}
These are again normalised proportions $\sum_{cc'} T^a_{cc'}=1$, but now they are so for each $a$, not for all together. This motivates its notation as a superscript to distinguish it from the normalised proportions of initial and final carriers. We can then express the effective transformation of a country as
\begin{equation}
    T_{cc'} = P_l T^l_{cc'} + \sum_a P_a T^a_{cc'} \, , 
\end{equation}
where $P_l = \sum_{cc'} Q_{cc'}$ is the total energy lost directly from carriers and $T^l_{cc'} = Q_{cc'} / P_l$ can be interpreted as the effective transformation of losses. The total proportion of the flows introduced or used by the activity (out of the total) can also be calculated by marginalising over rows or columns of the effective transformation of the activity
\begin{equation}
    P^{in, a}_{c} = \frac{\sum_{c'} P_{cc'a}}{P_a} = \sum_{c'} T^a_{cc'} ~~~~~~~~ P^{use, a}_{c'} = \sum_c T^a_{cc'} \, .
\end{equation}

Because JRC-ENERNET provides a detailed account of the activities and carriers in the energy systems of different countries, the effective transformation can vary significantly across them and even over time. However, we are mostly interested in broader properties, such as the total amount of energy introduced in the form of any non-renewable carrier, or the total use of renewables by transformation activities in general. For this, we can take advantage of the paths being mutually exclusive, and reason that the total flows where a type of node is specified instead of the node itself will simply be the sum over paths with nodes in that type. For example, the total amount of energy introduced as a carrier $c$ and used in a final form $c'$ by any activity in a particular activity group $\alpha$, such as final demand activities, is $P_{cc'\alpha} = \sum_{a \in \alpha} P_{cc'a}$. In combination with the total amount of flows ending at activities in the group $P_\alpha = \sum_{a \in \alpha} P_a$, we can define an effective transformation for activity groups instead of individual ones or the entire country,
\begin{equation}
    T^{\alpha}_{cc'} = \frac{\sum_{a \in \alpha} P_{cc'a}}{P_\alpha} \, .
\end{equation}
Note that the effective transformation of the country can also be expressed as an average of the activity group effective transformations, $T_{cc'} = P_l T^l_{cc'} + \sum_{\alpha} P_\alpha T^{\alpha}_{cc'}$.

Similarly, the total proportion of flows entering the system in the form of any carrier $c$ in a group $\kappa$, for example non-renewables, and leaving it in any form $c'$ in the group $\kappa'$, such as electricity, through a specific activity $a$ is
\begin{equation}
    P_{\kappa\kappa'a} = \sum_{\substack{c \in \kappa \\ c' \in \kappa'}} P_{cc'a} \, .
\end{equation}
This allows us to aggregate over initial and final carriers in different groups for any of the effective transformations (i.e. at the level of the entire country, individual activities, or groups of them) maintaining the same properties,
\begin{equation}\label{eq.typesET}
    T_{\kappa\kappa'} = \sum_{\substack{c \in \kappa \\ c' \in \kappa'}} T_{cc'} ~~~~~~~~ T^{\alpha}_{\kappa\kappa'} = \sum_{\substack{c \in \kappa \\ c' \in \kappa'}} T^{\alpha}_{cc'} ~~~~~~~~ T^a_{\kappa\kappa'} = \sum_{\substack{c \in \kappa \\ c' \in \kappa'}} T^a_{cc'} \, .
\end{equation}
We can also sum over rows or columns to obtain the total energy introduced or used (respectively) in the form of a particular carrier at any of these levels. These are also all normalised proportions, meaning that they account for all possible different paths at that scale.

However, this does not consider all possible paths in the entire country where energy is introduced in the form of $\kappa$ and leaves it in the form of $\kappa'$, which would represent the effective transformation of the entire country. This is because energy can also be lost directly from the carrier nodes in proportions $Q_{\kappa\kappa'}$ out of the total in the system. Thus, if we define $P_{l} = \sum_{\kappa\kappa'}Q_{\kappa\kappa'}$ and $T^{l}_{\kappa\kappa'} = Q_{\kappa\kappa'}/P_{l}$, the effective transformation of an entire country is
\begin{equation}
    T_{\kappa\kappa'} = Q_{\kappa\kappa'} + \sum_a P_{\kappa\kappa'a} = P_{l} T^{l}_{\kappa\kappa'} + \sum_a P_a T^a_{\kappa\kappa'} = P_{l} T^{l}_{\kappa\kappa'} + \sum_{\alpha} P_{\alpha} T^{\alpha}_{\kappa\kappa'}\, ,
    \label{eq:effective_transformation}
\end{equation} either in terms of the effective transformations of activities or groups of them. 

The effective transformations also provide a proxy for the progress of the energy transition at each resolution. For example, the proportion of flows in the entire country that are introduced in the form of non-renewables is the sum over elements of the non-renewable row in the effective transformation of the country. Similarly, the proportion of flows introduced in the form of non-renewables out of the total used by a particular activity or group of them is the sum over the same row of the respective effective transformation,
\begin{equation}   P^{in}_{NRN} = \sum_{\kappa'} T_{NRN,\kappa'} ~~~~~~~~ P^{in,\alpha}_{NRN} = \sum_{\kappa'} T^{\alpha}_{NRN,\kappa'} ~~~~~~~~ P^{in,a}_{NRN} = \sum_{\kappa'} T^a_{NRN,\kappa'} 
   \label{eq:input}
\end{equation}
Likewise, the proportion of flows in the entire country that leave it (end paths either by use or loss) in the form of non-renewables is the sum over the elements of the non-renewable column of the effective transformation of the country. This is also valid for the proportion of energy used (end paths) in the form non-renewables by a particular activity or group
\begin{equation}
    P^{use}_{NRN} = \sum_{\kappa} T_{\kappa,NRN} ~~~~~~~~ P^{use,\alpha}_{NRN} = \sum_{\kappa} T^{\alpha}_{\kappa,NRN} ~~~~~~~~ P^{use,a}_{NRN} = \sum_{\kappa} T^a_{\kappa,NRN} \, .
    \label{eq:use}
\end{equation}

\subsection*{Herfindahl index}

With the different characterisations of energy flows through the system through composing proportions, we are now ready to describe the heterogeneity of the system through these proportions and the Herfindahl index. The Herfindahl index is a common metric to evaluate the diversification of financial portfolios. If the $N$ assets in the portfolio are indexed $i \in \mathbf{N}_{\leq N}$, and their share of the total in the portfolio is $p_i$, then the Herfindahl index of the portfolio is 
\begin{equation}
    H = \sum_i {p_i}^2 \, .
\end{equation}

The Herfindahl index takes its maximum value when one of the shares $i_o$ holds the entire value of the portfolio, $p_{i_o} = 1$, and all the rest hold no share $p_{i \neq i_o} = 0$. The minimum value is achieved when all assets in the portfolio hold the same share $p_i = 1/N$. In each of these cases, the Herfindahl index case the values $1$ and $1/N$ respectively, and between them takes values that increase capturing how far the shares are from a uniform distribution (for a fixed number of shares $N$). This is a mathematical consequence of any collection of shares $1 \geq p_i \geq 0$ that collectively capture a total $\sum_i p_i = 1$, and therefore it can be extended to any collection of contributions to a total. Indeed, the Herfindahl index is related to the Renyi entropy, an information theoretic measure of how heterogeneous a distribution is.

We therefore apply the Herfindahl index to measure how diverse the contributions to flows in the energy system are according to different characterisations. We introduce only a minor change in the definition, which consists of subtracting the Herfindahl index from $1$ so that the value is a minimum of $0$ when all shares of the total are held by one value,
\begin{equation}
    H^* = 1 - \sum_i {p_i}^2 \, .
\end{equation}
This is applied to the shares of the total energy flows in the system represented in the effective transformations. Note that in this case each share $i$ is essentially a pair of carrier (types) either at the activity, group or country level
\begin{equation}
    H =1 - \sum_{\kappa\kappa'} {T_{\kappa\kappa'}}^2 ~~~~~~~~ H^{\alpha} = 1 - \sum_{\kappa\kappa'} {T^{\alpha}_{\kappa\kappa'}}^2 ~~~~~~~~ H^a = 1 - \sum_{\kappa\kappa'} {T^a_{\kappa\kappa'}}^2 \, .
\end{equation}
Note that the choice of calculating the Herfindahl index on the level of carrier types, instead of carriers, is a consequence of the reasoning of the Herfindahl index being applicable to distributions with the same number of shares $N$, while the carriers in a country over time or between countries can vary greatly.

\subsection*{Perspectives on the progress of the energy transition}
\label{sec:transition_progress}

\begin{figure}[ht!]
    \centering
    \includegraphics[width=0.7\linewidth]{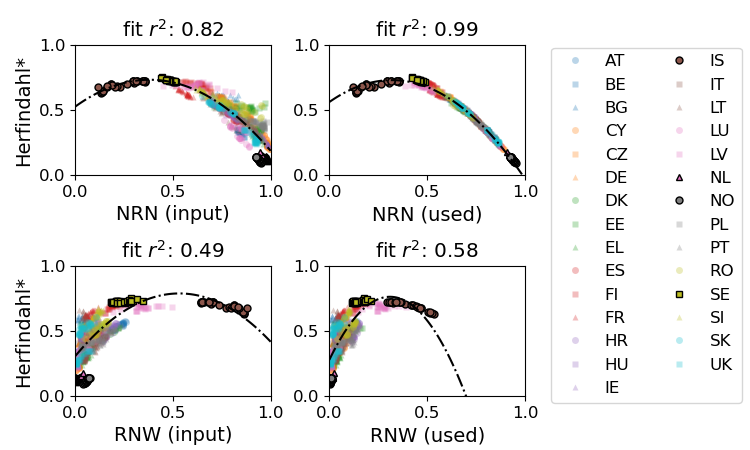}
    \caption{On the top, the Herfindahl index of the energy systems of all countries over time as a function of the proportion of non-renewables, input and used on the left and right sides respectively. This is the same data as shown in \cref{fig:type_effective_transformation_herf}). On the bottom, the Herfindahl index is plotted as a function of the proportion of renewables, also input and used on the left and right sides. Note that the transition unfolds in the opposite directions on the upper and lower figures. In the latter case, they show less agreement with the quadratic behaviour observed for the non-renewables, observed through the $r^2$ of the fit. This is essentially due to the fact that it is not renewables alone that replace non-renewables, but rather in combination with the other types, namely electricity and heat.}
    \label{fig:type_level_effective_transformation_herfindahl_NRNRNW}
\end{figure}

\begin{figure}[ht!]
    \centering
    \includegraphics[width=0.7\linewidth]{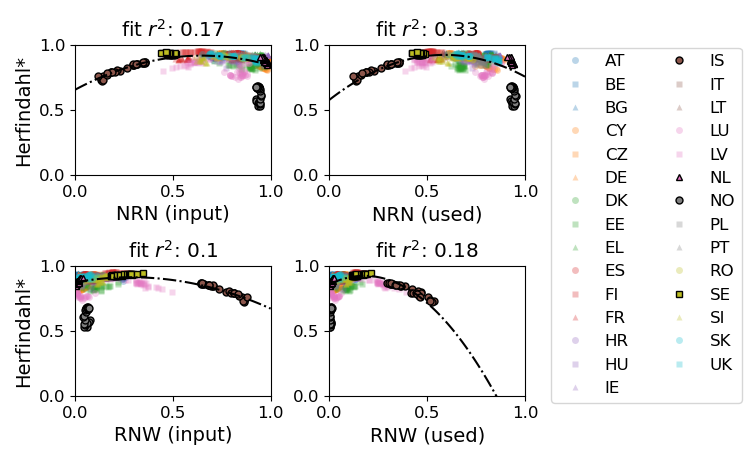}
    \caption{Herfindahl index measured by the node-level effective transformation instead of the type-level one. The $r^2$ for quadratic fits are much lower than for the type-level effective transformation, reflecting a larger disparity between countries due to having different number of carriers.}
    \label{fig:node_level_effective_transformation_herfindahl_NRNRNW}
\end{figure}

In this section we show that the qualitative behaviour of the Herfindahl index with respect to the progress of the transition does not change significantly with the level of aggregation of activities -node instead of group level (Figure \ref{fig:node_level_effective_transformation_herfindahl_NRNRNW}), nor with the perspective on the energy system -from the use of input perspective or the RNW on NRN shares (Fig \ref{fig:type_level_effective_transformation_herfindahl_NRNRNW}). Note that changing perspective or level of aggregation affects the fitting but not the shape.

\section*{Author Contributions}

Noam Abadi, Rossana Mastrandrea, Franco Ruzzenenti and Andrea Diaz  designed and shaped the analysis, and wrote and reviewed the manuscript. Noam Abadi and Rossana Mastrandrea carried out the main analysis. Noam Abadi, Rossana Mastrandrea and Andrea Diaz prepared the figures.

\section*{Acknowledgements}

We would like to thank Stephan Tchung-Ming, Herian Atma, Yang Wang and Jin Yan for fruitful discussions on the analysis of energy systems that helped guide this work. We also acknowledge the support of the Universities of Groningen and Turin, the IMT school for advanced sciences, and the Joint Research Centre of the European Commission. 

\section*{Funding}

This publication is part of the project ``ShaRepair: Sharing and repairing as circular consumption practices in everyday urban life'' (KICH1.ED07.22.011) which is (partly) financed by the Dutch Research Council (NWO). It also receives support from the PRIN 2022 project “The Role of the Public and Private Sectors in Pharmaceutical Breakthrough Innovations (3PBI)” (CUP2022S4EAS9).

\section*{Conflict of Interest} 

The authors declare no conflict of interest.

\bibliographystyle{unsrt}
\bibliography{ref}

\end{document}

\section*{Supplementary figures}

\begin{figure}[ht!]
    \centering
    \includegraphics[width=0.95\linewidth]{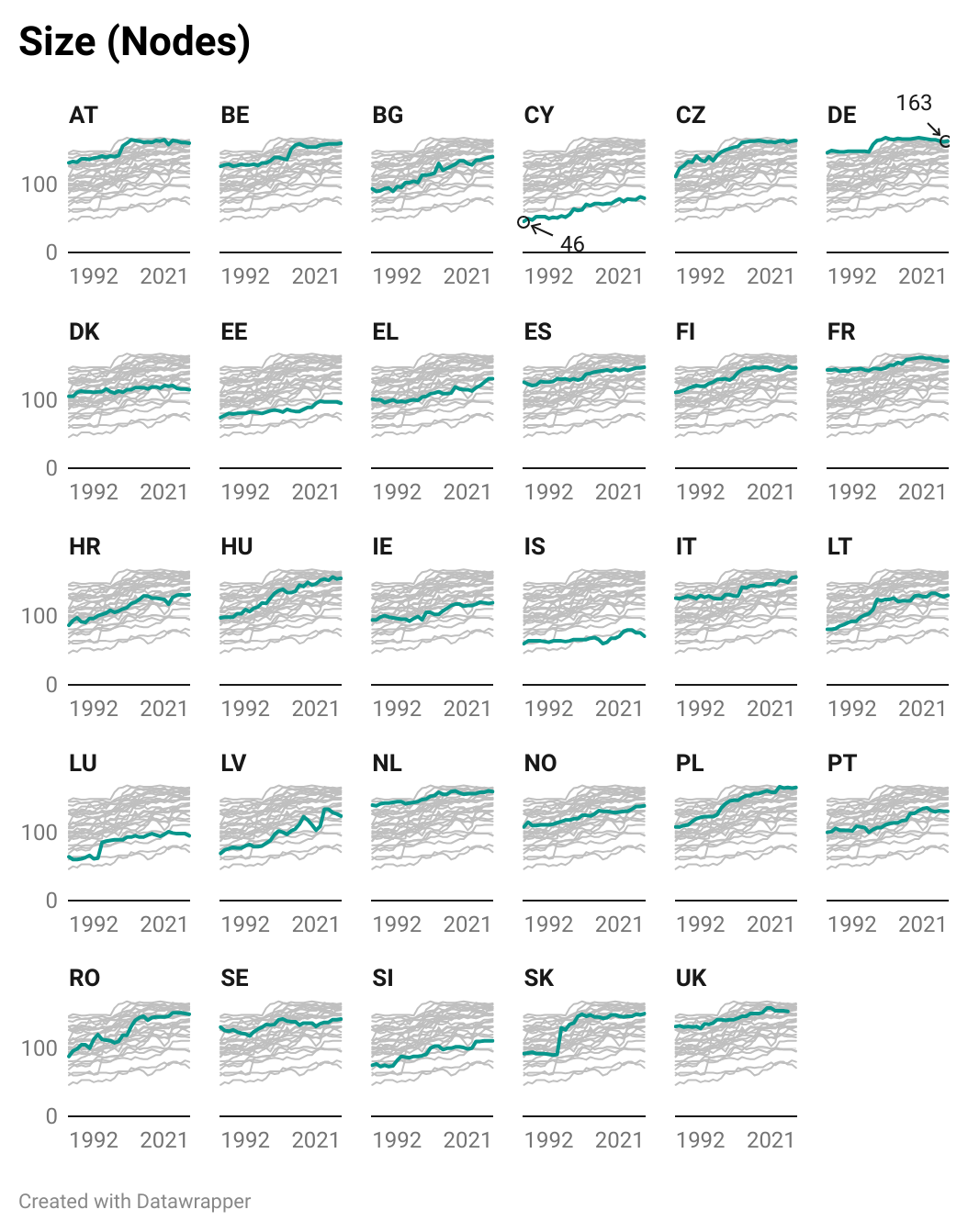}
    \caption{Number of nodes in different countries over time.}
    \label{fig:number_of_nodes}
\end{figure}

\begin{figure}[ht!]
    \centering
    \includegraphics[width=0.95\linewidth]{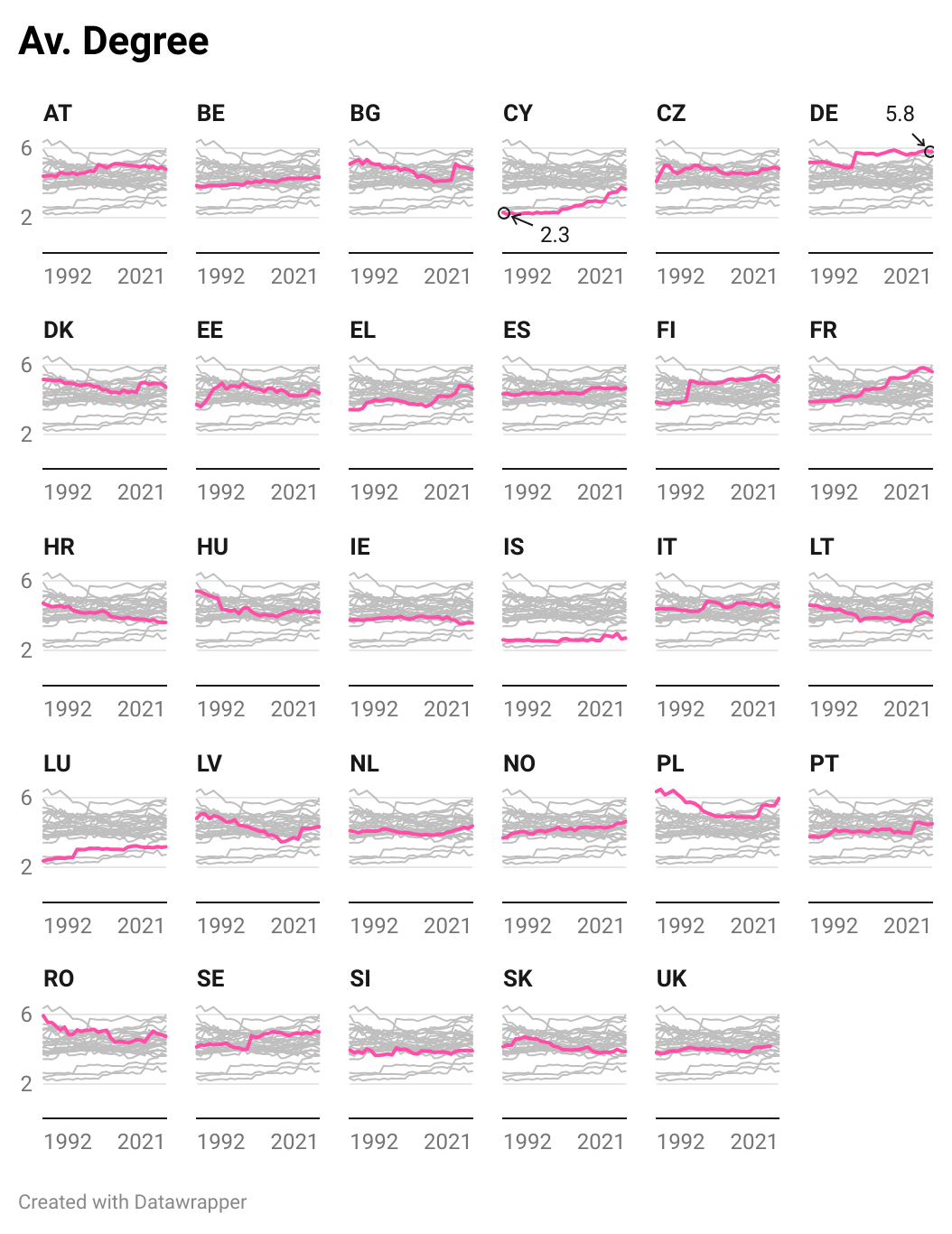}
    \caption{Average degree in different countries over time.}
    \label{fig:average_degree}
\end{figure}

\begin{figure}[ht!]
    \centering
    \includegraphics[width=\linewidth]{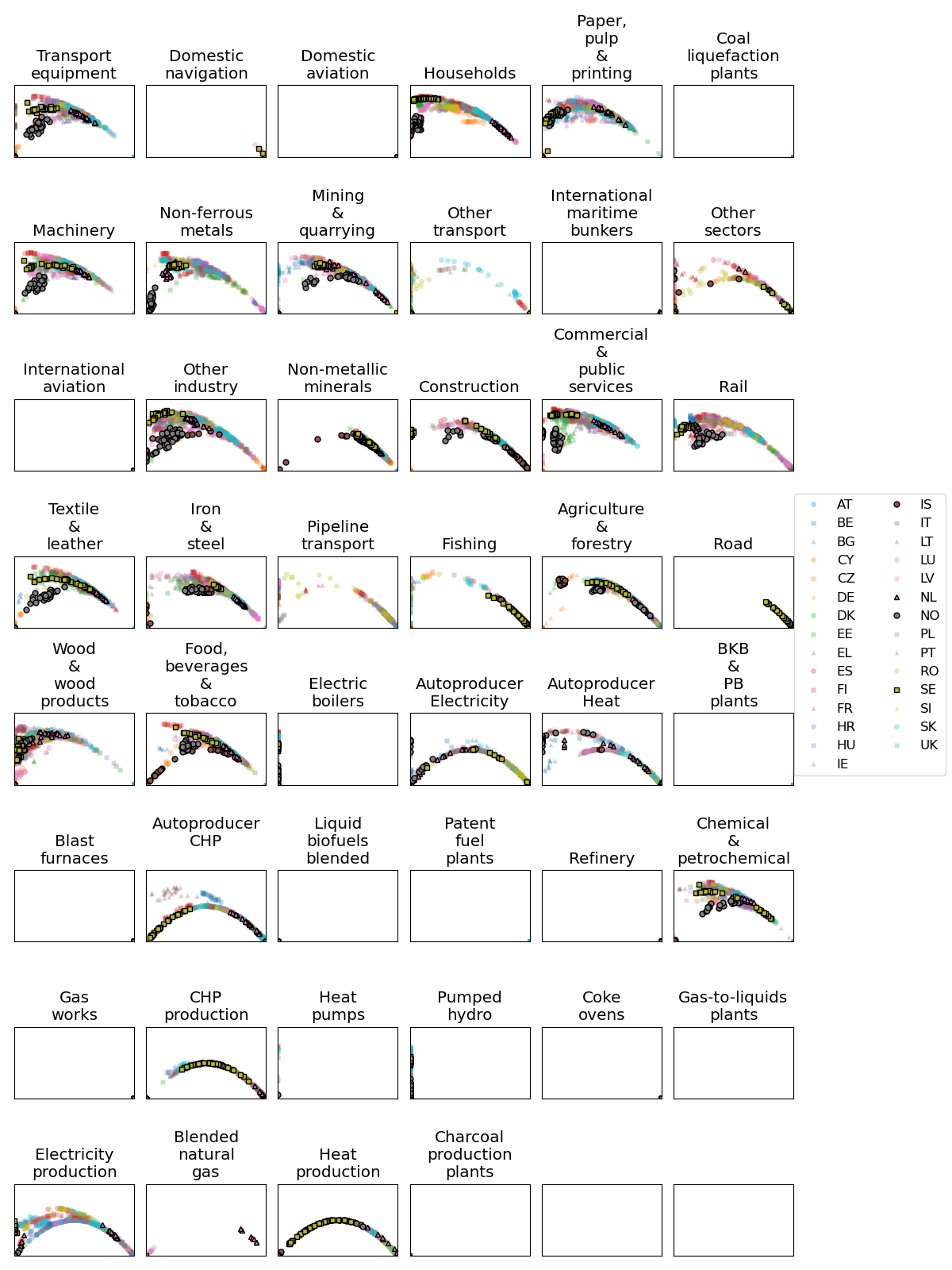}
    \caption{Herfindahl index as a function of proportion of NRN used by each activity for all countries.}
    \label{fig:activity_level_herfindahl}
\end{figure}